# Unfolding selection to infer individual risk heterogeneity for optimising disease forecasts and policy development

**Authors:** M. Gabriela M. Gomes,[1,2,*] Nicholas A. Feasey,[3,4] Marcelo U. Ferreira,[5,6] E. James LaCourse,[3] Kate E. Langwig,[7] Lisa Reimer,[3] Beate Ringwald,[3] Jamie Rylance,[3,4] J. Russell Stothard,[3] Miriam Taegtmeyer,[3] Dianne J. Terlouw,[3,4] Rachel Tolhurst,[3] Tom Wingfield,[3,8,9] Stephen B. Gordon.[3,4]

[1] Department of Mathematics and Statistics, University of Strathclyde, Glasgow, United Kingdom.

[2] Centro de Matemática e Aplicações, Faculdade de Ciências e Tecnologia, Universidade Nova de Lisboa, Caparica, Portugal.

[3] Liverpool School of Tropical Medicine, Liverpool, United Kingdom.

[4] Malawi Liverpool Wellcome Trust Programme of Clinical Tropical Research, Blantyre, Malawi.

[5] Institute of Biomedical Sciences, University of São Paulo, São Paulo, Brazil.

[6] Global Health and Tropical Medicine, Institute of Hygiene and Tropical Medicine, Nova University of Lisbon, Lisbon, Portugal.

[7] Department of Biological Sciences, Virginia Tech, Blacksburg, VA, USA.

[8] Tropical and Infectious Diseases Unit, Royal Liverpool University and Broadgreen Hospitals NHS Trust, Liverpool, United Kingdom.

[9] Social Medicine, Infectious Diseases and Migration Group, Department of Public Health Sciences, Karolinska Institutet, Stockholm, Sweden.

[*] Correspondence to: gabriela.gomes@strath.ac.uk.

**Key points:**

Models with incomplete heterogeneity overpredict infection burdens and overestimate intervention impacts.

Individual heterogeneity can be inferred holistically by estimating how much selection occurs as susceptible subpopulations are depleted through infection.

These methods rely on *unfolding selection gradients*.

**Abstract:** Mathematical models are increasing adopted for setting targets for disease prevention and control. As model-informed policies are implemented, however, the inaccuracies of some forecasts become apparent, for example overprediction of infection burdens and overestimation of intervention impacts. Here, we attribute these discrepancies to methodological limitations in capturing the heterogeneities of real-world systems. The mechanisms underpinning single factors for infection and their interactions determine individual propensities to acquire disease. These are potentially so numerous that to attain a full mechanistic description may be unfeasible. To contribute constructively to the development of health policies, model developers either leave factors out (reductionism) or adopt a broader but coarse description (holism). In our view, predictive capacity requires holistic descriptions of heterogeneity which are currently underutilised in infectious disease epidemiology but common in other disciplines.

Setting realistic targets and developing feasible strategies for disease prevention and control depends on representative models. These can be conceptual, experimental, or mathematical. Mathematical modelling was established in infectious diseases over a century ago [1-3]. Propelled by the discovery of aetiological agents for infectious diseases, and Koch's postulates, models have focused on the complexities of pathogen transmission and evolution to understand and predict disease trends in greater depth [4]. This has led to their adoption by decision makers to inform national and international policy. However, as model-informed policies are being implemented, the inaccuracies of some forecasts are increasingly apparent, most notably their tendency to overpredict infection burdens and overestimate the impact of control interventions [5-7]. Here, we discuss how these discrepancies could be explained by methodological limitations in capturing the effects of individual variation in real-world systems. We suggest improvements that derive from theory developed in demography to study frailty variation [8]. Using simulations, we illustrate the problem by incorporating individual variation within infectious diseases models and formulate a pragmatic approach to estimate the most impactful forms of heterogeneity.

**Heterogeneity affects accuracy of model forecasts**

We use the examples of acquired immunodeficiency syndrome (AIDS) and coronavirus disease 2019 (COVID-19) to illustrate the effects that individual heterogeneity can have on the performance of mathematical models for the dynamics of endemic and epidemic diseases.

*Endemic infectious diseases*

Since the detection of AIDS in the early 1980s, it has been evident that heterogeneity in individual sexual behaviours needed to be considered in mathematical models for the transmission of the causative agent – the Human Immunodeficiency Virus (HIV) [9]. Much research has been devoted to measuring contact networks in diverse settings and by different methods, to attempt to reproduce transmission dynamics accurately [10-12]. However, other equally important sources of inter-individual variation were overlooked. For example, unmodelled heterogeneity in infectiousness and susceptibility led to over-emphasis of the acute-phase HIV as a driver of new infections [13]. This resulting in an overlooked opportunity in "treatment as prevention" measures.

The problem of unaccounted heterogeneity in disease forecast models can be illustrated with the simplest mathematical description of infectious disease transmission in a host population. Figure 1 shows the prevalence of infection over time under three alternative scenarios: all individuals are at equal risk of acquiring infection (black trajectories); individual risk is affected by a factor that modifies either their susceptibility to infection (blue) or exposure through connectivity with other individuals (green). Homogeneous models assign every individual a risk factor of 1 (black frequency plot), whereas heterogeneous risk derives from a distribution with mean one (blue and green density plots). As the virus spreads within the population, individuals at higher risk are predominantly infected as indicated at endemic equilibrium (Figure 1 A, B, C, density plots on the right, coloured red) and after 100 years of control (Figure 1 D, E, F). The control strategy applied to endemic equilibrium in the figure is the 90-90-90 treatment as prevention target advocated by the Joint United Nations Programme on HIV/AIDS (UNAIDS) whereby 90% of HIV-infected individuals should be detected, with 90% of these receiving antiretroviral therapy, and 90% of these should achieve viral suppression (becoming effectively non-infectious).

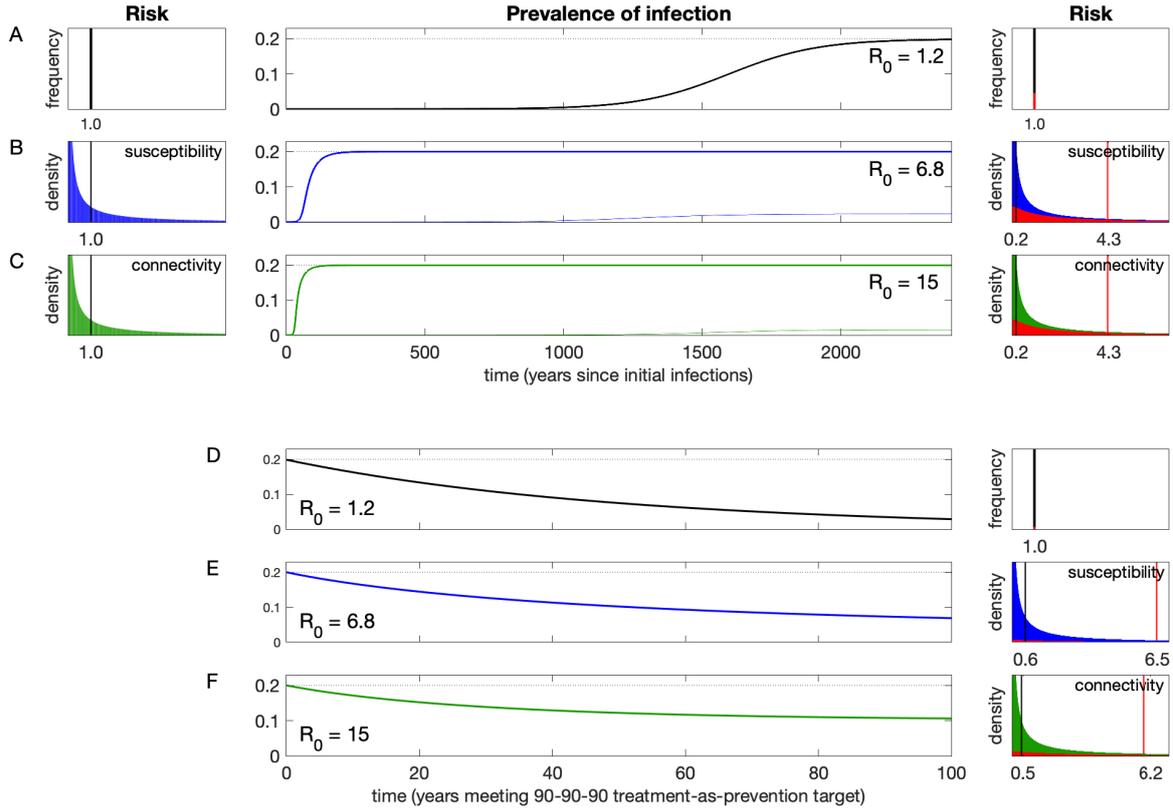

**Figure 1: Prevalence trajectories under homogeneous and heterogeneous models.** Risk distributions are simulated in three scenarios: homogeneous (A, D) [notice the unrealistic time scale in A]; distributed susceptibility to infection with variance 10 (B, E); distributed connectivity with variance 10 (C, F). In disease-free equilibrium, individuals differ in potential risk in scenarios B and C, but not in scenario A (risk panels on the left). The vertical lines mark the mean risk values (1 in all cases). At endemic equilibrium, individuals with higher risk are predominantly infected (risk panels on the right, where red vertical lines mark mean baseline risk among individuals who eventually became infected), resulting in reduced mean risk among those who remain uninfected (black vertical lines). To compensate for this selection effect, heterogeneous models require a higher $R_0$ to attain the same endemic prevalence (A, B, C). Interventions that reduce infection also reduce selection pressure, which unintendedly increases mean risk in the uninfected subpopulation and undesirably reduces intervention impact (D, E, F). Models: homogeneous (A, D) $dS/dt = \mu - \beta IS - \mu S$, $dI/dt = \beta IS - \mu I$, and $R_0 = \beta/\mu$; heterogeneous susceptibility (B, E) $dS(x)/dt = q(x)\mu - \beta \int I(u) du \, xS(x) - \mu S(x)$, $dI(x)/dt = \beta \int I(u) du \, xS(x) - \mu I(x)$, and $R_0 = \beta/\mu$; heterogeneous connectivity (C, F) $dS(x)/dt = q(x)\mu - \beta \int uI(u) du \, xS(x) - \mu S(x)$, $dI(x)/dt = \beta \int uI(u) du \, xS(x) - \mu I(x)$, and $R_0 = \int u^2 q(u) du \, \beta/\mu$. In heterogeneous models, $q(x)$ is a probability density function with mean 1 and variance 10. Gamma distributions were used for concreteness.

Figure 1 shows that heterogeneous models that account for wide biological and social variation require higher basic reproduction numbers ($R_0$) to reach a given endemic level and predict less impact for control efforts when compared with the homogeneous counterpart model. This holds true regardless of whether heterogeneity affects susceptibility or connectivity and is extensive to more realistic combinations of the two traits. At endemic equilibrium, individuals at higher risk are predominantly infected (red distributions have mean greater than one as marked by the red vertical lines), and hence those who remain uninfected are individuals with lower risk (blue and green distributions have mean lower than one as marked by the black vertical lines). Thus, the mean risk in the uninfected but susceptible subpopulation decreases, and the epidemic decelerates (thin blue and green curves); higher values of $R_0$ are consequently required if the heterogeneous models are to attain the same endemic level as the homogeneous formulation (heavy blue and green curves). Finally, interventions are less impactful under heterogeneity because $R_0$ is implicitly higher. Indeed, these biases could help explain trends in HIV incidence data which lag substantially behind targets informed by model predictions, even in settings that have reached the 90-90-90 implementation targets [5,6].

*Epidemic infectious diseases*

At the end of 2019, a novel severe acute respiratory syndrome coronavirus (SARS-CoV-2) isolated from a patient in China began to spread worldwide causing the COVID-19 pandemic. Countrywide epidemics have been extensively analysed and modelled throughout the world. Initial studies projected attack rates of around 90% if transmission had been left unmitigated [14], while subsequent reports noted that individual variation in susceptibility or exposure might flatten epidemic curves and reduce these estimates substantially [15-17], as shown in Figure 2 (compare the blue [heterogeneous susceptibility] and green [heterogeneous connectivity] curves with the black [homogeneous]). Moreover, these types of variation that are subject to selection by the force of infection tend to affect population measures of risk ratios leading to biased interpretations if realistic heterogeneity is not accounted for. For example, the bottom panel in Figure 2 illustrates how reinfection risk is likely to be overestimated when heterogeneity is neglected (black horizontal line represents individual risk ratio while blue and green curves depict time-dependent population risk ratios under heterogeneous susceptibility and connectivity, respectively).

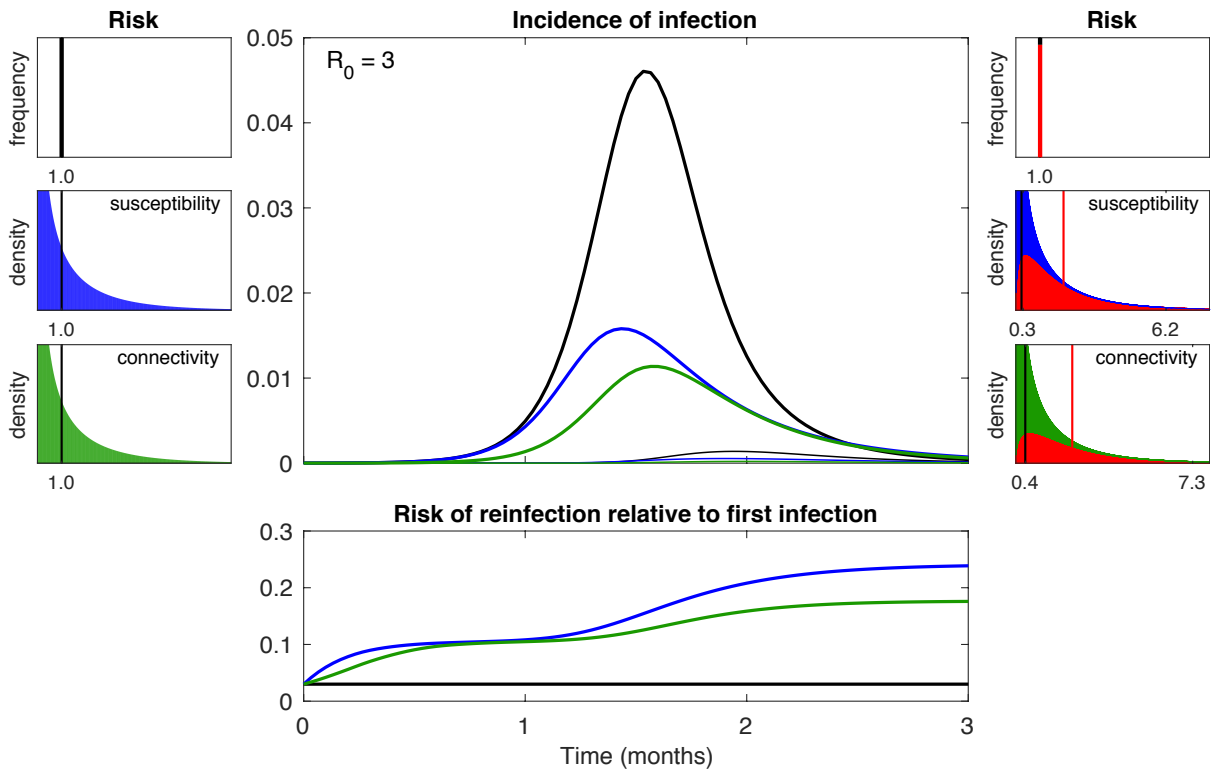

**Figure 2: Incidence trajectories under homogeneous and heterogeneous models.** Risk distributions are simulated in three scenarios: homogeneous (black); distributed susceptibility to infection with variance 2.5 (blue); distributed connectivity with variance 2.5 (green). On the main panel, heavy lines represent first infection and thin lines are reinfection. Left panels represent distributions of potential individual risk prior to the outbreak, with vertical lines marking mean risk values (1 in all cases). As the epidemic progresses, individuals with higher risk are predominantly infected, depleting the susceptible pool in a selective manner and decelerating epidemic growth. Right panels show in red the risk distributions among individuals who have been infected over 3 months of epidemic spread (mean greater than one when risk is heterogeneous, as marked by red vertical lines) and the reduced mean risk among those who have not been affected (black vertical lines). Models: homogeneous (black) $dS/dt = -\beta IS$, $dI/dt = \beta I(S + \sigma R) - \gamma I$, $dR/dt = \gamma I - \sigma \beta IR$, and $R_0 = \beta/\gamma$;

heterogeneous susceptibility (blue) $dS(x)/dt = -\beta \int I(u)du\, xS(x)$, $dI(x)/dt = \beta \int I(u)du\, x[S(x) + \sigma R(x)] - \gamma I(x)$, $dR(x)/dt = \gamma I(x) - \sigma\beta \int I(u)du\, xR(x)$, and $R_0 = \beta/\gamma$; heterogeneous connectivity (green) $dS(x)/dt = -\beta \int uI(u)du\, xS(x)$, $dI(x)/dt = \beta \int uI(u)du\, x[S(x) + \sigma R(x)] - \gamma I(x)$, $dR(x)/dt = \gamma I(x) - \sigma\beta \int uI(u)du\, xR(x)$ and $R_0 = \int u^2 q(u)du\, \beta/\gamma$. In heterogeneous models, $q(x)$ is a probability density function with mean 1 and variance 2.5. Gamma distributions were used for concreteness. Parameter $\sigma$ represents the risk of reinfection of each individual relative to their own risk of first infection, here assumed $\sigma = 0.03$. The bottom panel depicts the average risk of reinfection (over the subpopulation at risk of reinfection) relative to the average risk of first infection (over the subpopulation at risk of first infection).

Representing individual variation is necessary to predict infectious disease dynamics and inform policy. Epidemic curves for COVID-19 are widely available, and it is possible to construct models with inbuilt risk distributions. Their shapes can be inferred by assessing their ability to mould simulated trajectories to observed epidemics, while accounting for realistic social and biomedical interventions [17].

Variation in infectiousness has been critical to the occurrence of explosive outbreaks resulting from superspreaders in both 2002 SARS-CoV-1 and 2019 SARS-CoV-2 [18,19]. This heterogeneity is different, however: variation in infectiousness does not lead to selective depletion of the susceptible pool as variation in susceptibility or connectivity do, i.e., models with and without variation in infectiousness perform identically when implemented deterministically and only differ through stochasticity processes.

The need to account for heterogeneity in risk of acquiring infections is generally applicable across other models of infectious disease epidemiology. Moreover, similar issues arise in methods intended to evaluate the efficacy of interventions from experimental studies.

**Heterogeneity and vaccine efficacy over time and across settings**

Individual variation in susceptibility or exposure to infection induces biases in cohort studies and clinical trials. Vaccine efficacy trials offer a useful illustration of the problem and give insight into a potential solution. In a vaccine trial, two groups of individuals are randomised to receive a vaccine or placebo and disease occurrences are recorded in each group. As disease affects predominantly higher-risk individuals, the mean risk among those who remain unaffected decreases and disease incidence declines. In the vaccine group the same trend will occur at a slower pace (presuming that the vaccine protects to some degree). As a result, the two randomised groups become different over time with more highly susceptible individuals remaining in the vaccine group. The vaccine efficacy, described as a ratio of cases in vaccinated compared to control group, therefore appears to wane (Figure 3) [20,21]. This effect will be stronger in settings where transmission intensity is higher, inducing a trend of seemingly declining efficacy with disease burden [22]. The concept is illustrated in Figure 3 by simulating a vaccine trial with heterogeneous and homogeneous models analogous to those utilised in Figures 1 and 2.

Selection on individual variation in disease susceptibility thus offers an explanation for vaccine efficacy trends that is entirely based on population level heterogeneity, in contrast with individual waning of vaccine-induced immunity [23]. It is important to disentangle their roles, as both may occur concurrently in a trial and lead to different interpretations of the same data. For example, waning of individual vaccine-induced immunity may superficially look the same as a population decline due to selection on individual variation. To capture this in a timely manner requires multicentre trial designs with sites carefully selected over a gradient of transmission intensities (e.g., optimally spaced along the incidence axis in Figure 3 C, F), and analyses performed by fitting curves generated by models that incorporate individual variation. An alternative and more tightly controlled approach would be to use experimental designs in human infection challenge studies where these are available [24] to generate dose-response curves and apply similar models [25]. These approaches have recently been successfully tested in animal systems [26-28].

An essential purpose in suggesting these study designs (randomised controlled trials with long follow-up, multicentre trials over a gradient of transmission intensities, or dose-response infection challenges) is to enable the unfolding of selection gradients in such a way that individual risk heterogeneity can be inferred from observed patterns of infection.

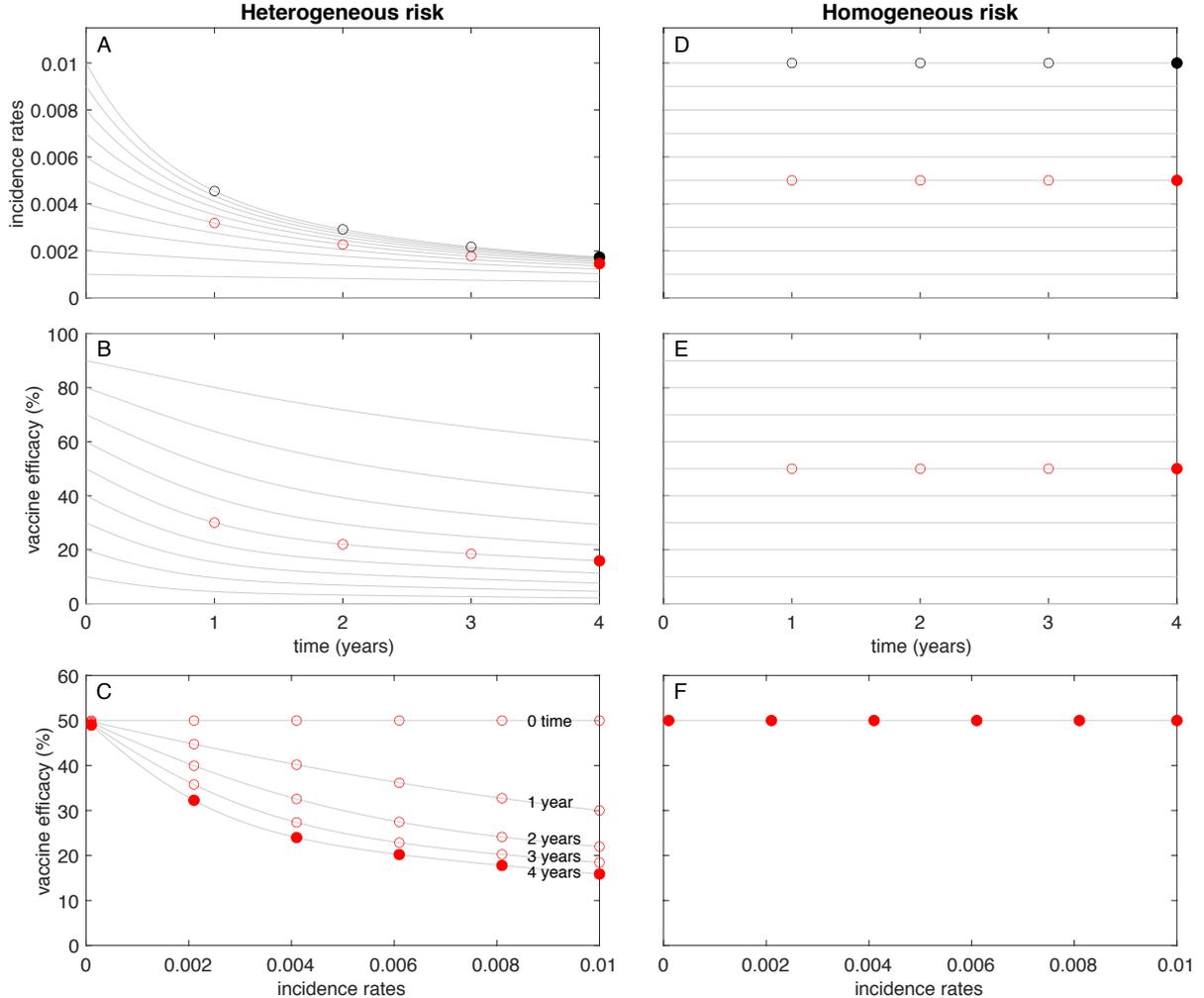

**Figure 3: Vaccine efficacy trajectories under homogeneous and heterogeneous models.** A, B, C, Heterogeneous susceptibility or exposure (gamma-distributed with mean 1 and variance 10); D, E, F, Homogeneous model. Models: (homogeneous) $dS_c/dt = -\lambda S_c$, $dI_c/dt = \lambda S_c$, and $dS_v/dt = -\sigma\lambda S_v$, $dI_v/dt = \sigma\lambda S_v$; (heterogeneous) $dS_c(x)/dt = -\lambda x S_c(x)$, $dI_c(x)/dt = \lambda x S_c(x)$, and $dS_v(x)/dt = -\sigma\lambda x S_v(x)$, $dI_v(x)/dt = \sigma\lambda x S_v(x)$. Vaccine efficacy is calculated as $[1 - r_v(t)/r_c(t)] \times 100$, where $r_v$ and $r_c$ represent the incidences in vaccinated ($v$) and control ($c$) groups, respectively: (homogeneous) $r_c(t) = \lambda$; $r_v(t) = \sigma\lambda$; (heterogeneous) $r_c(t) = \lambda \int x S_c(x,t)dx / \int S_c(x,t)dx$; $r_v(t) = \sigma\lambda \int x S_v(x,t)dx / \int S_v(x,t)dx$.

**Inferring risk heterogeneities by unfolding selection gradients**

Heterogeneities in predisposition to infection depend on the mode of transmission. In respiratory infections, heterogeneity may arise from variation in exposure of the susceptible host to the pathogen, or the competence of host immune systems to control it. These two processes have multiple component factors. Some of the most studied are age, patterns of inter-personal contacts, exposure to smoke, nutritional status, pre-existing respiratory illness such as asthma or chronic obstructive pulmonary disease, and the presence of other concomitant diseases such as diabetes and HIV. Enteric diseases have other heterogeneities determined by the source and dose of contaminated sources. Vector-borne pathogens may be transmitted by mosquitoes, ticks, snails, and other intermediate hosts, where the risk of onward transmission is affected by heterogeneities in exposure and susceptibility across a complex range of host, demographic, environmental and social factors. As for sexually transmitted disease specific factors include to behaviour, age, gender, and sexual orientation.

The mechanisms underpinning single factors for infection and their interactions determine individual propensities to acquire disease. These factors are potentially so numerous and intertwined that to attain a full mechanistic description is likely unfeasible. Even if a list of all putative factors were available, the measurement of effect sizes might be subject to selection within cohorts resulting in underestimated variances [29]. To contribute constructively to the development of health policies, model building involves compromises between leaving factors out (reductionism) or adopting a broader but coarse description (holism). Holistic descriptions of heterogeneity are currently underutilised in infectious diseases.

Descriptive measures of individual variation can be formulated into disease transmission models, whether they depict endemic [30,31] or epidemic [17] processes, in much the same way that they are used to describe risk inequality in non-communicable diseases, such as cancer [32], or non-health disciplines, such as economics, offering a holistic approach to improve the predictive capacity of models. Having conceived the model, the challenge becomes the quantification of relevant statistical dispersion parameters. In epidemic diseases, characterised by marked temporal dynamics, individual variation can be most simply estimated by fitting dynamic models to series of reported infections, hospitalisations, or deaths [17]. As for endemic diseases, typically these do not display as much change over time that we might learn from, so we need to be more creative at unfolding selection gradients. This may involve stratifying the population into groups of individuals with similar risk, which may be as granular as individual level for frequent diseases, such as influenza or malaria [31], geographical units for diseases which cluster by proximity, such as tuberculosis [30], or familial relatedness when there is a clear genetic contribution to risk, such as cancer [32]. By recording disease events in each group, specific incidence rates can be calculated and ranked. The formulated models, which incorporate explicit distributions of individual risk, are then fitted to the stratified data to estimate the extent of individual variation among other parameters of interest. Once developed, these models will automatically adjust average risks in susceptible subpopulations to changes in transmission intensity, should these occur naturally or in response to interventions. Not subject to the selection biases described in this paper, this modelling approach inherently enables more accurate impact forecasts for use in policy development.

**Conclusion**

There is compelling evidence for the utility of holistic indicators that account for individual variation in disease risk, admitting that heterogeneity is so vast in real-world systems that complete mechanistic reconstructions may be unachievable. Inspired by other population disciplines and supported by successful applications in both infectious and non-communicable diseases, we describe methods of study design and analyses that enable holistic inferences of heterogeneity by estimating how much selection occurs as susceptible subpopulations are depleted through infection. These methods rely on *unfolding selection gradients*. Applying these approaches to epidemiology offers significant advantages: disease models could provide more accurate descriptions of intervention effects, and better disease forecasts.

**Contributions**

MGMG conceived the idea and all authors contributed to the development and writing of this article.

**Declaration of interests**

We declare no competing interests.

**Acknowledgements**

TW is supported by grants from the Wellcome Trust, UK (209075/Z/17/Z) and the Medical Research Council, Department for International Development, and Wellcome Trust (Joint Global Health Trials, MR/V004832/1).